\documentclass[aps,physrev,reprint,groupedaddress]{revtex4-2}

\usepackage{amsmath} 
\usepackage{subcaption}
\usepackage{graphicx} 
\usepackage{xcolor}
\usepackage[colorlinks=true,
            pdfborder={0 0 0},
            citecolor=blue,
            linkcolor=blue,
            urlcolor=blue]{hyperref}
            
\begin{document}


\title{Chaos Generation and Control with Molecular Optomechanical System}



\author{Hui-Hui Xu$^{1}$}
\author{Jian-Zhuang Wu$^{2}$}\altaffiliation{These authors contributed equally to this work.}
\author{Lei-Bo Deng$^{1}$} 
\author{E Wu$^{1}$}\email{towue@163.com}
\author{Yong-Hong Ma$^{1}$}\email{myh_dlut@126.com}

\affiliation{$^{1}$School of Science, Inner Mongolia University
of Science and Technology, Baotou 014010, China}
\affiliation{$^{2}$Center for Quantum Sciences and School of Physics, Northeast Normal University, Changchun 130117, China}


\date{\today}

\begin{abstract}
    Chaos is central to secure communication and physical random-number generation. Conventional cavity-optomechanical implementations, however, usually rely on weak single-photon optomechanical coupling and low-frequency mechanical modes, so access to deterministic chaotic dynamics often requires large driving power and careful suppression of thermal noise. Here we theoretically study a hybrid molecular optomechanical system formed by coupling a plasmonic nanocavity to a whispering-gallery-mode (WGM) microcavity. The plasmonic nanocavity provides terahertz-scale single-photon optomechanical coupling to a molecular vibration, while the WGM resonator offers a low-loss photonic channel that mitigates the short plasmon lifetime. By integrating the semiclassical equations of motion and evaluating the largest Lyapunov exponent, we map the nonlinear dynamical regimes in the parameter spaces of WGM detuning, plasmon--WGM coupling, and plasmon--vibration optomechanical coupling. We show that increasing the plasmon--vibration coupling drives the system from self-sustained oscillations to chaos through a period-doubling cascade. At moderate coupling strengths, isolated chaos windows can be opened or closed by tuning the WGM detuning and the inter-cavity coupling. These results identify molecular optomechanics as a controllable room-temperature platform for on-chip chaotic light generation and random-signal applications.
\end{abstract}

\maketitle

\section{Introduction}\label{one}

Chaos, characterized by extreme sensitivity to initial conditions, is a cornerstone of nonlinear science and holds significant promise for secure communication and encryption \cite{AllariaSynchronization01,WOS:000380760000001,WOS:000245871200063,thompson2002nonlinear}. Recent breakthroughs include using chaotic dynamics to suppress decoherence, terahertz wideband chaotic masking communication, and ultrahigh-speed physical random number generators with rates exceeding 100\,Gbps \cite{ChenPhysical25,PengBroadband24}. Although chaos has been observed in diverse platforms such as cavity optomechanics \cite{LUPT15,MonifiOptomechanically16,Navarro-UrriosNonlinear17,ZhangNonreciprocal21,RoqueNonlinear20,ZhangSqueezing-induced25,ZhuCavity23}, cavity magnomechanics \cite{LiuPhase-mediated19,LiNonlinear23,PengCavity24,WangMagnon19}, and optoelectromechanical systems \cite{Xuchaos25,WangControllable16}, and the typical period-doubling bifurcation route has been identified \cite{BakemeierRoute15,LiRoutes24,HalefRoute25}, conventional cavity optomechanical implementations rely on micro- and nanomechanical resonators with single-photon optomechanical coupling strengths only at the hertz to kilohertz level and mechanical frequencies in the megahertz to gigahertz range. As a result, reaching the nonlinear threshold for chaos demands high driving power, and the low-frequency mechanical modes are plagued by abundant thermal phonons at room temperature, with thermal noise severely degrading the coherence and deterministic nature of the chaos~\cite{GenesRobust08,WangReservoir-Engineered13}. Consequently, experimental observation of conventional optomechanical chaos often requires cryogenic environments or complex external modulation, and achieving low-threshold, highly controllable chaos at room temperature remains an open challenge.

The convergence of non-Hermitian physics and plasmonics has recently opened new avenues for stronger optomechanical coupling \cite{LiBoosting24,EstebanMolecular22,MaTwo-Polariton26} and richer nonlinear effects~\cite{RoelliMolecular16,SchmidtQuantum16,SchmidtLinking17,LombardiPulsed18}. Exploiting the subwavelength field confinement of plasmonic nanocavities, the single-photon optomechanical coupling strength can be boosted to the terahertz level, several orders of magnitude higher than in conventional microcavity systems \cite{RoelliMolecular16,AndersonTwo-Color18,BellCoherent25,KimUltrasmall16}. Simultaneously, the intrinsic frequencies of molecular vibrational modes are typically tens of terahertz, giving a thermal phonon occupation at room temperature as low as $\sim 0.01$, i.e., essentially in the quantum ground state, thereby providing an extremely clean dynamical environment. However, the intrinsically high optical losses of plasmonic modes severely limit the intracavity photon lifetime and the build-up of coherence. Coupling a high-quality-factor WGM optical microcavity to a plasmonic nanocavity perfectly compensates for this drawback: through evanescent coupling, Stokes photons generated in the plasmonic cavity are efficiently transferred to the low-loss WGM resonator, thereby suppressing the decoherence induced by plasmonic dissipation. This hybrid architecture not only enables robust room-temperature photon–phonon entanglement \cite{HuangCollective24,YinMolecular-Optomechanical26,BerinyuyQuantum26, Loirette-PelousAddressing26,BarzanjehStationary19} but also opens new possibilities for exploring strong nonlinear dynamics \cite{BenzSingle-molecule16, BaumbergPicocavities22,BaumbergExtreme19,TangRobust26,YinMolecular25,BellCoherent25}. Despite extensive studies of chaos in cavity optomechanics, the conditions for chaos generation, control laws, and multidimensional parameter phase diagrams in such hybrid plasmonic–WGM systems incorporating molecular vibrational modes have not yet been systematically revealed.

Here, we theoretically investigate chaotic dynamics in this hybrid molecular optomechanical system. By solving the semiclassical equations of motion and computing the largest Lyapunov exponent, we systematically map the nonlinear dynamics in the parameter spaces spanned by the plasmon-vibration optomechanical coupling strength $g_c$, the plasmon–WGM coupling strength $J$, and the optical detuning $\Delta_a$. We uncover a complete period-doubling route to chaos and demonstrate that $g_c$ acts as the primary nonlinear driving force, while $J$ and $\Delta_a$ provide highly efficient control over the chaotic regime. Remarkably, even at moderate $g_c$, isolated chaos windows can be opened and tuned solely by adjusting the optical detuning and the inter-cavity coupling, underscoring the high degree of optical controllability. These findings establish molecular optomechanical systems as a versatile platform for room-temperature, on-chip chaotic light sources and random signal generators.

The organization of this paper is outlined below. In Section \ref{two}, we establish the system model and Hamiltonian formulation, followed by the derivation of the dynamical evolution equations. Section \ref{three} presents a systematic parametric study of chaotic dynamics, employing Lyapunov exponent analysis in conjunction with multiple complementary chaos characterization approaches. Finally, Section \ref{four} recapitulates the principal contributions of this research and elaborates on their broader significance.

\section{Model and Hamiltonian}\label{two}

\begin{widetext}

    \begin{figure}[h]
    \centering
    \includegraphics[width=0.9\linewidth]{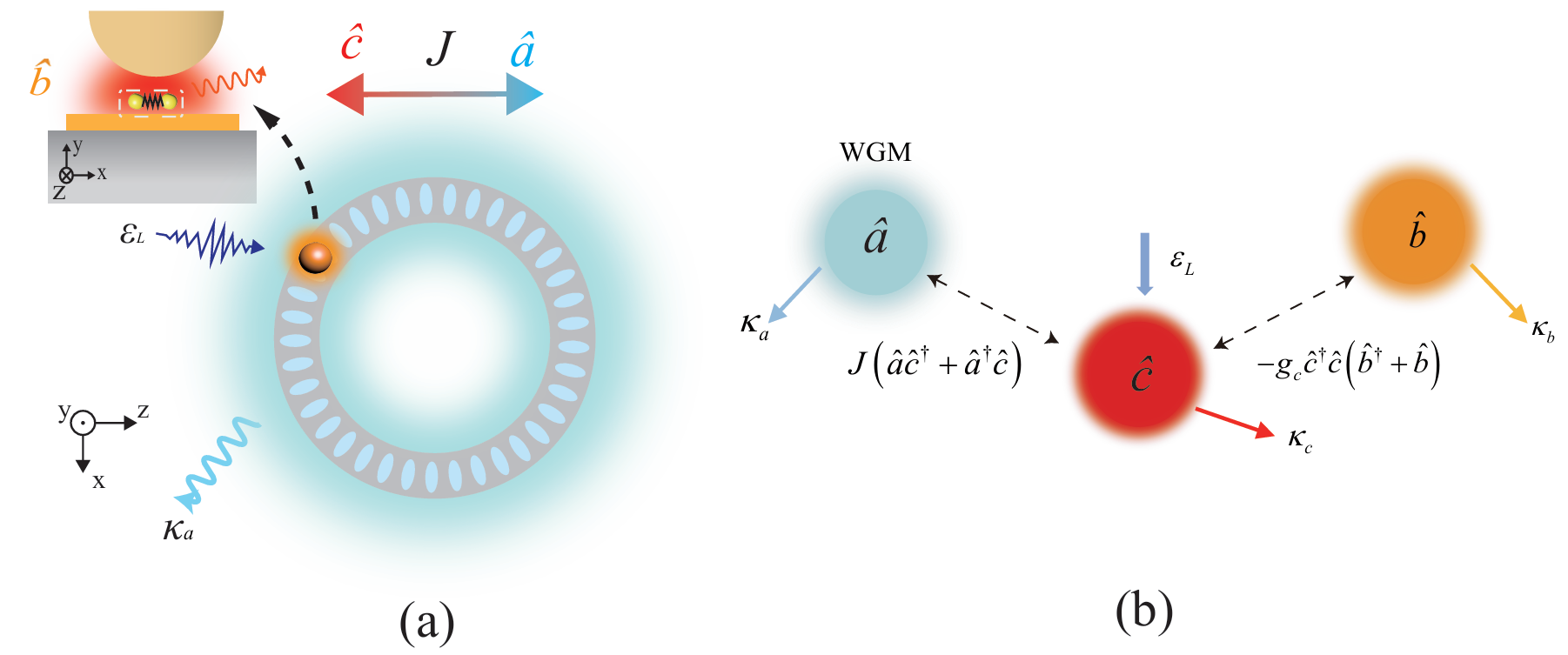}
    \caption{Schematic of the hybrid molecular cavity optomechanical system. (a) The hybrid platform consists of a WGM resonator and a nanoparticle-on-mirror (NPoM) plasmonic nanocavity containing a single molecule. (b) Equivalent schematic of the three-mode interaction. The WGM mode $\hat{a}$ (frequency $\omega_a$, decay rate $\kappa_a$) is coupled to the plasmonic mode $\hat{c}$ ($\omega_c$, $\kappa_c$) via an evanescent wave with coupling strength $J$. The plasmonic mode interacts with the molecular vibrational mode $\hat{b}$ ($\omega_b$, $\kappa_b$) through radiation pressure, with coupling strength $g_c$.}
    \label{model1}
\end{figure}

\end{widetext}

The system consists of a plasmonic nanocavity coupled to a WGM resonator (Fig.~\ref{model1}(a)). The ``nanoparticle-on-mirror'' (NPoM) plasmonic cavity containing a biphenyl-4-thiol molecule supports both a plasmonic mode ($\omega_c$, $\kappa_c$) and a molecular vibrational mode ($\omega_b$, $\kappa_b$). The WGM resonator hosts an optical mode ($\omega_a$, $\kappa_a$), which is coupled to the plasmonic mode via an evanescent wave with strength $J$; $\kappa_a$ includes the intrinsic cavity loss and the additional dissipation introduced by the gold mirror (Fig.~\ref{model1}(b)). The plasmon–molecule interaction is mediated by radiation pressure, with coupling strength $g_c$. This architecture leverages the broadband subwavelength confinement of the plasmonic structure to enhance $g_c$, while utilizing the low-loss $\kappa_a$ of the WGM to preserve quantum coherence. When the plasmonic mode is coherently driven with power $P$ at frequency $\omega_L$, the effective Hamiltonian of the system in the rotating frame is
\begin{align}
    \begin{aligned}
    \frac{\hat{H}}{\hbar} =& \Delta_a \hat{a}^{\dagger}\hat{a} + \Delta_c \hat{c}^{\dagger}\hat{c} + \omega_b \hat{b}^{\dagger}\hat{b} - g_c \hat{c}^{\dagger}\hat{c}(\hat{b}^{\dagger} + \hat{b})  \\  &+ J(\hat{a}\hat{c}^{\dagger} + \hat{a}^{\dagger}\hat{c}) + i\epsilon_L(\hat{c}^{\dagger} - \hat{c}),
\end{aligned}
\end{align}
where \(\hat{a}(\hat{a}^{\dagger})\), \(\hat{c}(\hat{c}^{\dagger})\), and \(\hat{b}(\hat{b}^{\dagger})\) are the annihilation (creation) operators for the WGM optical mode, the plasmonic mode, and the molecular vibrational mode, respectively. \(\Delta_{a,c} = \omega_{a,c} - \omega_L\) are the detunings between each mode and the driving laser. The third term is the free molecular vibrational energy. The fourth term describes the radiation-pressure-type optomechanical coupling, where the plasmon number density modulates the vibrational displacement, with strength \(g_c\). The fifth term is the beam-splitter-type evanescent coupling between the WGM mode and the plasmonic mode, with strength $J$. The last term represents the coherent drive of the plasmonic mode, with the driving amplitude \(\epsilon_L = \sqrt{2\eta\kappa_c P/\hbar\omega_L}\), where \(P\) is the driving power and \(\eta\) is the radiation efficiency.

In the semiclassical limit, neglecting quantum correlations and replacing operators with their expectation values, the Heisenberg–Langevin equations reduce to
\begin{align}
    \begin{aligned}
\dot{a} &= -(i\Delta_a + \kappa_a)a - iJc, \\
\dot{b} &= -(i\omega_b + \kappa_b)b + ig_c|c|^2, \\
\dot{c} &= -(i\Delta_c + \kappa_c)c + ig_c c(b^* + b) - iJa + \epsilon_L.
\end{aligned}\label{banLWfun}
\end{align}

To facilitate stability analysis and Lyapunov exponent calculation, the complex amplitudes are decomposed into real and imaginary parts: \(a = a_r + ia_i\), \(b = b_r + ib_i\), \(c = c_r + ic_i\). Substituting these into Eq.~(\ref{banLWfun}) yields a six-dimensional autonomous nonlinear dynamical system
\begin{align}
    \begin{aligned}
\dot{a}_r &= -\kappa_a a_r + \Delta_a a_i + Jc_i, \\
\dot{a}_i &= -\kappa_a a_i - \Delta_a a_r - Jc_r, \\
\dot{b}_r &= -\kappa_b b_r + \omega_b b_i, \\
\dot{b}_i &= -\kappa_b b_i - \omega_b b_r + g_c(c_r^2 + c_i^2), \\
\dot{c}_r &= -\kappa_c c_r + \Delta_c c_i + J a_i - 2g_c b_r c_i + \epsilon_L, \\
\dot{c}_i &= -\kappa_c c_i - \Delta_c c_r - J a_r + 2g_c b_r c_r.
\end{aligned}\label{shixvfun}
\end{align}
The nonlinearities \(g_c(c_r^2 + c_i^2)\) and \(2g_c b_r c_{i,r}\) originate from the radiation-pressure interaction and are the physical origin of complex dynamical behaviors such as chaos. Meanwhile, the beam-splitter interaction \(J\) linearly couples the WGM mode and the plasmonic mode, correlating their dynamics and further enriching the system's behavior. Since the system has more than three degrees of freedom and the equations of motion are nonlinear, the basic conditions for chaotic motion are satisfied \cite{thompson2002nonlinear}. To characterize chaotic dynamics, we introduce the perturbation evolution of adjacent trajectories \(\delta o = (\delta a_{r},\delta a_{i},\delta b_{r},\delta b_{i},\delta c_{r},\delta c_{i} )^{T}\). The perturbation evolution is governed by the linearized equation \(\dot{\delta o} = M\delta o\), where the Jacobian matrix $M$ is

\begin{equation}
    \mathbf{M} = 
\begin{pmatrix}
-\kappa_a & \Delta_a & 0 & 0 & 0 & J \\
-\Delta_a & -\kappa_a & 0 & 0 & -J & 0 \\
0 & 0 & -\kappa_b & \omega_b & 0 & 0 \\
0 & 0 & -\omega_b & -\kappa_b & 2g_c c_r & 2g_c c_i \\
0 & J & -2g_c c_i & 0 & -\kappa_c & \Delta_c - 2g_c b_r \\
-J & 0 & 2g_c c_r & 0 & -\Delta_c + 2g_c b_r & -\kappa_c
\end{pmatrix}\label{yakebi}
\end{equation}
The matrix $M$ governs the divergence of nearby trajectories, serving as a key criterion for the onset of chaos and determining the stability properties of the system.

\section{Results and Discussion}\label{three}

We numerically integrate Eqs.~(\ref{shixvfun}) using a fourth-order Runge–Kutta algorithm and compute the largest Lyapunov exponent \(\lambda_{\max}\) via the Benettin algorithm, which evolves the reference trajectory and tangent vectors simultaneously with periodic Gram–Schmidt orthonormalization. \(\lambda_{\max} > 0\) signifies chaotic motion, \(\lambda_{\max} = 0\) corresponds to critical points such as period-doubling bifurcations, and \(\lambda_{\max} < 0\) indicates steady states or limit-cycle oscillations \cite{NowoczynUniversal26,escot2020brief,benettin1980lyapunov,datseris2018dynamicalsystems,baier1991chaotic}. Unless otherwise specified, the experimentally feasible parameters adopted in this work are as follows \cite{YuStrong26,BenzSingle-molecule16,JakobGiant23,BellCoherent25,LombardiPulsed18,KimUltrasmall16}: \(\omega_a/2\pi = \omega_c/2\pi = 370\) THz, \(\kappa_a/2\pi = 2.4\) GHz, \(\kappa_c/2\pi = 15\) THz, \(\omega_b/2\pi = 30\) THz, \(\kappa_b/2\pi = 0.01\) THz, \(g_c/2\pi \in [0, 10]\) THz, \(J/2\pi \in [0, 1.5]\) THz, the radiation efficiency is \(\eta = 0.45\), and the driving power is \(P=3.308~\mathrm{\mu W}\).

\subsection{Control by the plasmon-vibration coupling and optical detuning}

\begin{figure}[htbp]
    \centering
    \includegraphics[width=1\linewidth]{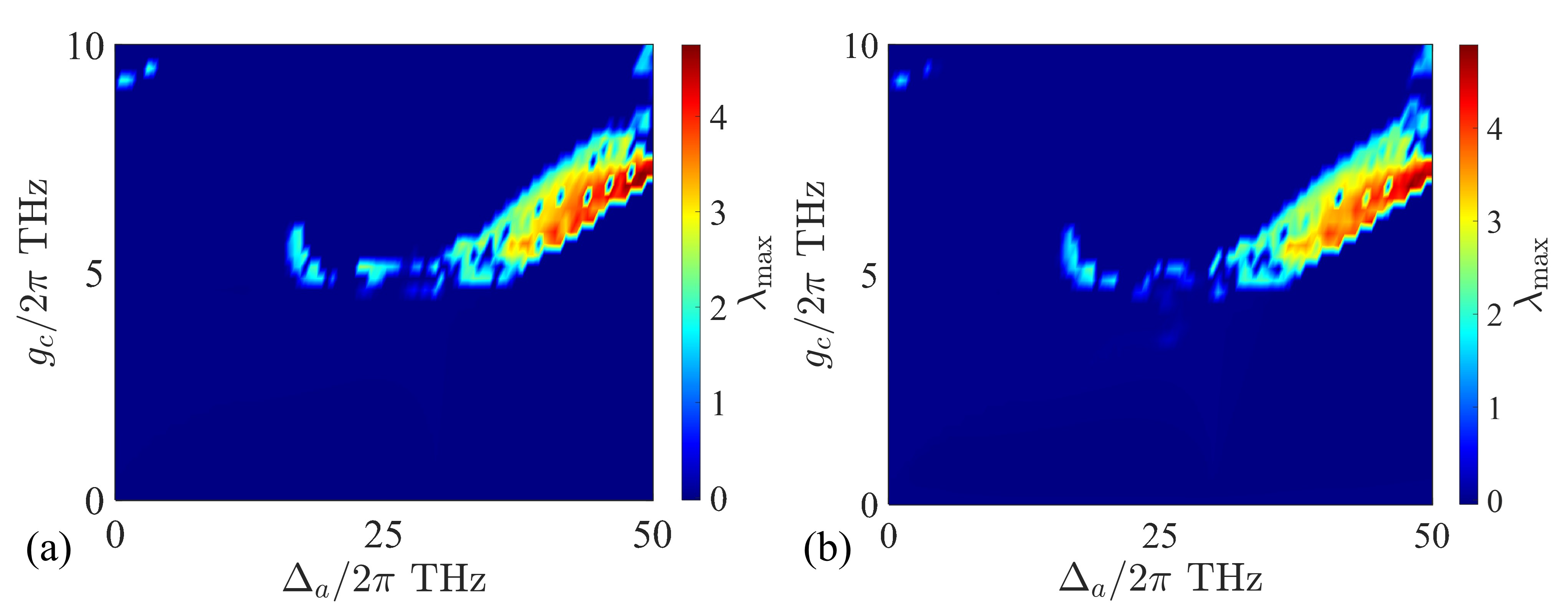}
    \caption{Phase diagrams of the largest Lyapunov exponent \(\lambda_{\max}\) in the plane of the WGM detuning \(\Delta_a\) and the plasmon-vibration coupling strength \(g_c\). (a) Plasmon–WGM coupling strength \(J/2\pi = 0.38\,\mathrm{THz}\); (b) \(J/2\pi = 0.70\,\mathrm{THz}\). The color bar represents the value of \(\lambda_{\max}\): warm colors (\(\lambda_{\max}>0\)) correspond to chaotic motion, and cold colors correspond to regular steady-state or periodic motion.}
    \label{fig:LE_Delta_a_gc_Jc0.38-0.7}
\end{figure}

Figure~\ref{fig:LE_Delta_a_gc_Jc0.38-0.7} shows the distribution of \(\lambda_{\max}\) in the parameter space spanned by the WGM detuning \(\Delta_a\) and the plasmon-vibration coupling strength \(g_c\). Over a wide detuning range, particularly for positive detuning \(\Delta_a > 0\), increasing \(g_c\) drives the system into chaos, and the chaotic region as well as the magnitude of \(\lambda_{\max}\) expand monotonically with \(g_c\). Physically, positive detuning redshifts the effective frequency of the plasmonic mode, bringing it closer to resonance with the molecular vibrational mode and significantly enhancing the energy exchange efficiency; negative detuning suppresses the nonlinear coupling. Hence chaos preferentially emerges for \(\Delta_a > 0\). The coupling \(g_c\) directly determines the strength of the nonlinear terms \(g_c(c_r^2 + c_i^2)\) and \(2g_c b_r c_{i,r}\): a larger plasmon–molecule coupling increases the mechanical displacement per photon, and the mutual driving between the plasmon intensity and the mechanical displacement intensifies. When this nonlinear feedback overcomes the intrinsic dissipation, the originally stable limit cycle collapses into a chaotic attractor via a period-doubling bifurcation cascade. In the weak-coupling regime, \(\lambda_{\max} \approx 0\) across the entire detuning range, and the system mainly exhibits regular steady states or self-sustained oscillations. This reflects a key advantage of the molecular platform: at room temperature the thermal phonon occupation of the vibrational mode is extremely low (\(\bar{n} \approx 0.01\)), allowing the system to maintain highly regular dynamics even under weak coupling, without being overwhelmed by thermal noise. Comparing Fig.~\ref{fig:LE_Delta_a_gc_Jc0.38-0.7}(a) (\(J = 0.38\) THz) and Fig.~\ref{fig:LE_Delta_a_gc_Jc0.38-0.7}(b) (\(J = 0.7\) THz) reveals that increasing the inter-cavity coupling strength \(J\) has only a minor influence on the chaos phase diagram.

\subsection{Cooperative Control by the Plasmon–WGM Coupling Strength and the plasmon-vibration coupling}

\begin{widetext}

\begin{figure}[t]
    \centering
    \includegraphics[width=1\linewidth]{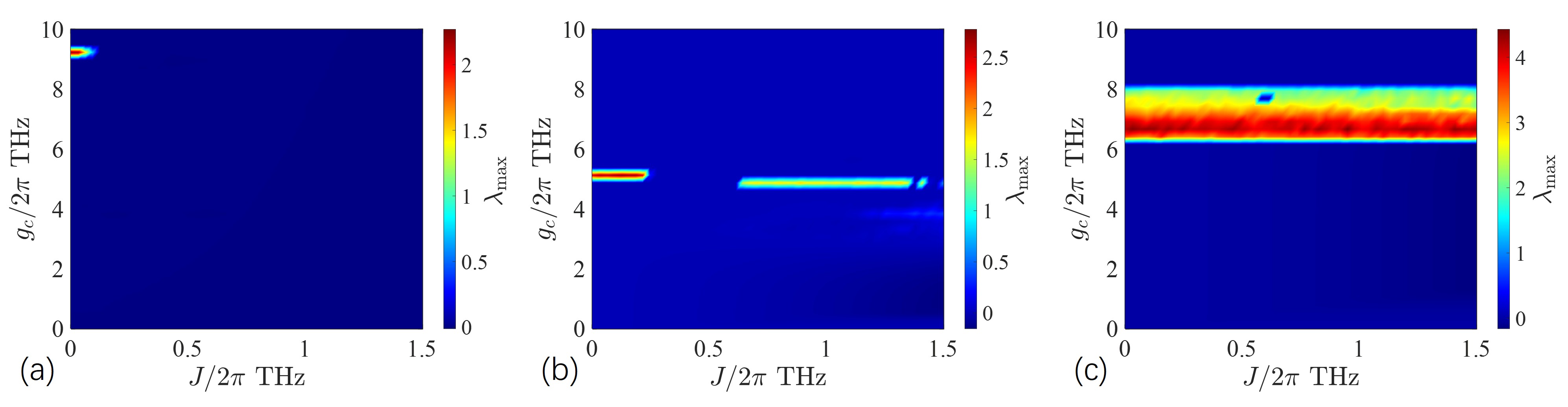}
    \caption{Phase diagrams of the largest Lyapunov exponent \(\lambda_{\max}\) in the plane of the plasmon–WGM coupling strength \(J\) and the plasmon-vibration coupling strength \(g_c\) under different WGM detunings. (a) \(\Delta_a/2\pi = 0\,\mathrm{THz}\); (b) \(\Delta_a/2\pi = 21\,\mathrm{THz}\); (c) \(\Delta_a/2\pi = 45\,\mathrm{THz}\).}
    \label{fig:LE_Jc_gc-Delta_a0-21-45}
\end{figure}

\end{widetext}

To further elucidate the cooperative control by \(J\) and \(g_c\), we plot phase diagrams of \(\lambda_{\max}\) in the \((J, g_c)\) plane for three representative WGM detunings (Fig.~\ref{fig:LE_Jc_gc-Delta_a0-21-45}). At zero detuning (Fig.~\ref{fig:LE_Jc_gc-Delta_a0-21-45}(a)), chaos appears only in a narrow region of high \(g_c\) (\(g_c/2\pi \approx 9~\mathrm{THz}\)) and very small \(J\) (\(J/2\pi\approx0.1\) THz); over a broad range of parameters the system remains in regular motion. This indicates that when both optical modes are resonant with the drive, the energy exchange remains largely linear, making it difficult to excite nonlinear chaotic behavior. Under intermediate detuning (\(\Delta_a/2\pi=21\) THz, Fig.~\ref{fig:LE_Jc_gc-Delta_a0-21-45}(b)), the effective optomechanical coupling is resonantly enhanced, yielding a narrow chaotic band. As \(J\) increases, the formation of supermodes temporarily suppresses the direct plasmon–molecule interaction, causing chaos to vanish; further increasing \(J\) reintroduces effective nonlinearity through mode hybridization and chaos reappears. When the detuning is increased to \(45\,\mathrm{THz}\) (Fig.~\ref{fig:LE_Jc_gc-Delta_a0-21-45}(c)), the chaotic region expands dramatically, covering almost the entire range \(g_c/2\pi = [6,8]\,\mathrm{THz}\), and the control effect of \(J\) weakens, with \(g_c\) becoming the dominant chaos parameter. These results clearly demonstrate the crucial role of optical detuning in chaos control: by tuning the detuning, the emergence and disappearance of chaos can be flexibly manipulated.

\subsection{Optical tuning of chaotic windows}
\begin{widetext}

\begin{figure}[h]
    \centering
    \includegraphics[width=0.9\linewidth]{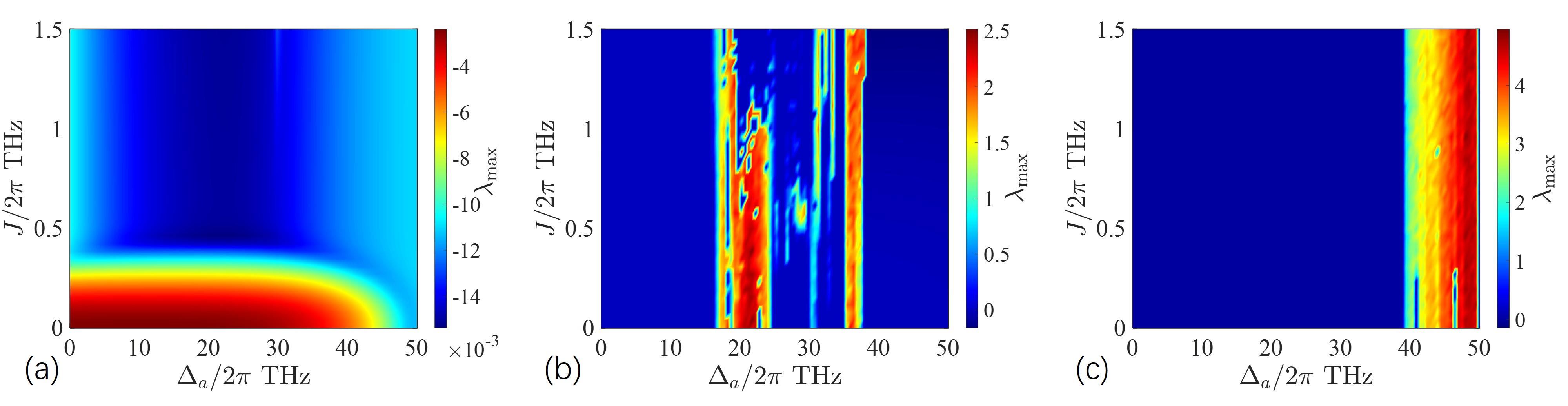}
    \caption{Phase diagrams of the largest Lyapunov exponent \(\lambda_{\max}\) as a function of the WGM detuning \(\Delta_a\) and the plasmon–WGM coupling strength \(J\) under different plasmon–vibration coupling strengths. (a) \(g_c/2\pi = 0.1~\mathrm{THz}\); (b) \(g_c/2\pi = 5~\mathrm{THz}\); (c) \(g_c/2\pi = 7~\mathrm{THz}\).}
    \label{fig:LE_Delta_a_Jc-gc0.1-5-7}
\end{figure}
    
\end{widetext}

When the plasmon-vibration coupling is extremely weak (\(g_c/2\pi=0.1~\mathrm{THz}\), Fig.~\ref{fig:LE_Delta_a_Jc-gc0.1-5-7}(a)), the system remains in regular motion (\(\lambda_{\max}<0\)) regardless of \(\Delta_a\) and \(J\), confirming that under weak coupling the energy exchange is nearly linear and cannot trigger chaos. As \(g_c/2\pi\) increases to \(5~\mathrm{THz}\) and \(7~\mathrm{THz}\) (Figs.~\ref{fig:LE_Delta_a_Jc-gc0.1-5-7}(b) and (c)), the chaotic region gradually expands, evolving from isolated chaotic bands into a continuous distribution, further confirming that enhancing \(g_c\) promotes chaos. Notably, at \(g_c/2\pi=5~\mathrm{THz}\), \(\Delta_a\) and \(J\) exert comparable control over chaos, and the system alternates between regular and chaotic motion as parameters vary, exhibiting the typical period-doubling route to chaos (see Fig.~\ref{fig:fenchaJ-0.38-gc5-7}). At \(g_c/2\pi=7~\mathrm{THz}\), the influence of \(J\) is significantly reduced; here \(g_c\) and \(\Delta_a\) dominate, and the chaos intensity is stronger. These results demonstrate that at moderate \(g_c\) the weights of \(\Delta_a\) and \(J\) are comparable, whereas in the deep nonlinear regime \(\Delta_a\) becomes the dominant control parameter and \(J\) provides only fine tuning of the chaos intensity. This hybrid molecular cavity optomechanical system thus offers a high degree of optical controllability: for suitable \(g_c\), one can switch between regular and chaotic states simply by precisely adjusting \(\Delta_a\) and \(J\), paving the way toward on-chip programmable chaotic light sources.

\subsection{Bifurcation route to chaos}

\begin{widetext}

    \begin{figure}[h]
    \centering
    \includegraphics[width=0.9\linewidth]{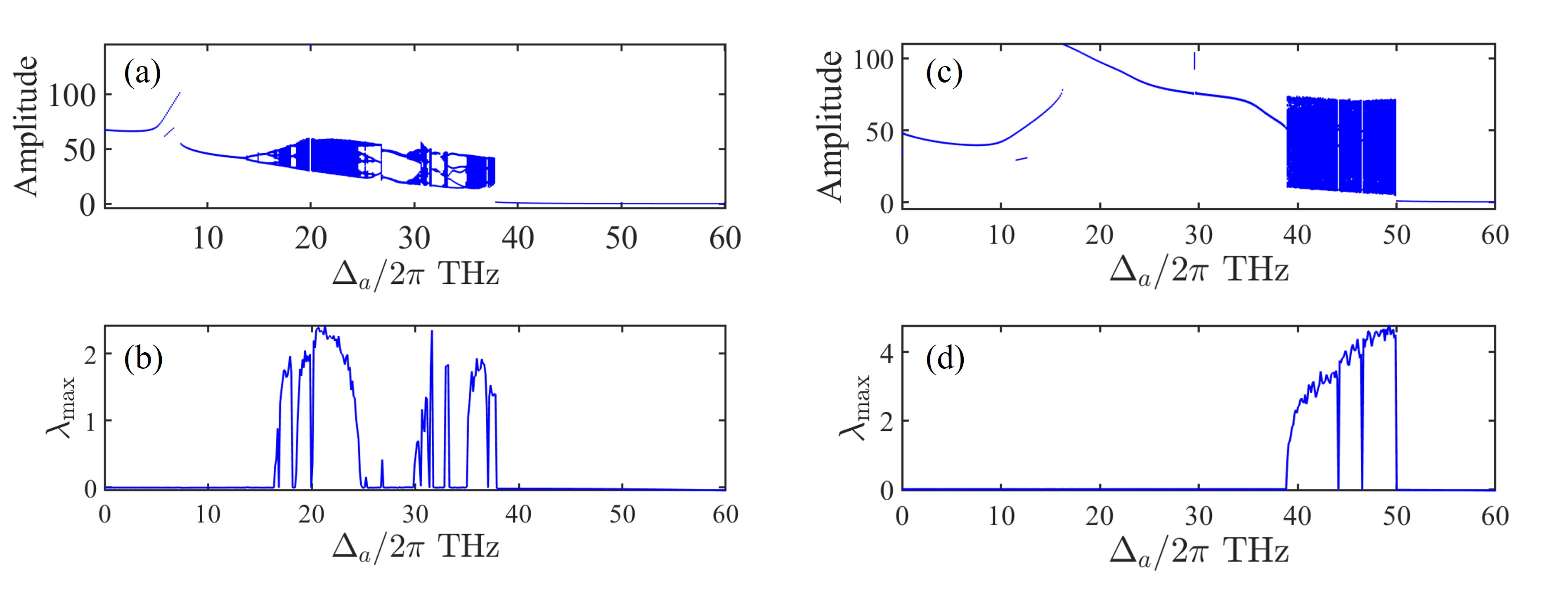}
    \caption{Bifurcation behavior at $J/2\pi=0.38\,\mathrm{THz}$. (a),(b) $g_c/2\pi=5\,\mathrm{THz}$: the upper panel shows the bifurcation diagram of the vibrational intensity $I_b=|b|^2$, and the lower panel shows the corresponding largest Lyapunov exponent $\lambda_{\max}$. (c),(d) Same as (a),(b), but for $g_c/2\pi=7\,\mathrm{THz}$.}
    \label{fig:fenchaJ-0.38-gc5-7}
\end{figure}

\end{widetext}

To further reveal the route to chaos, we plot the bifurcation diagram of the molecular vibration intensity \(I_b\) and the corresponding largest Lyapunov exponent as a function of the WGM detuning \(\Delta_a\) (Fig.~\ref{fig:fenchaJ-0.38-gc5-7}).

At \(J/2\pi=0.38~\mathrm{THz}\), for \(g_c/2\pi = 5~\mathrm{THz}\) (Figs.~\ref{fig:fenchaJ-0.38-gc5-7}(a)(b)), the bifurcation diagram around \(\Delta_a/2\pi \approx 18\) THz and \(>21\) THz exhibits a clear period-doubling cascade: period-1 \(\to\) period-2 \(\to\) period-4 \(\to\) chaos window, with \(\lambda_{\max}\) transitioning from negative through zero to positive. This explicitly verifies that the system enters chaos via the period-doubling bifurcation route. At this moderate coupling, the nonlinearity is not yet sufficient to completely destabilize the limit cycle; chaos appears only in isolated detuning windows where specific phase-matching conditions enhance the effective nonlinear driving force enough to overcome dissipation. When \(g_c/2\pi\) is increased to \(7~\mathrm{THz}\) (Figs.~\ref{fig:fenchaJ-0.38-gc5-7}(c)(d)), the chaotic region expands dramatically: the bifurcation points fill into continuous vertical bands over a wider detuning range, and \(\lambda_{\max}\) remains distinctly positive. This comparison clearly demonstrates that increasing \(g_c\) directly strengthens the core nonlinear terms, causing the regular self-sustained oscillation to become unstable through a cascade of period-doubling bifurcations and eventually form a chaotic attractor. The plasmon-vibration coupling strength \(g_c\) is thus identified as the core driving force for chaos. Physically, when \(g_c\) is small, the cavity frequency shift caused by the mechanical displacement is tiny and the optomechanical feedback is approximately linear; once \(g_c\) exceeds a threshold, the optically induced frequency shift becomes comparable to the cavity linewidth, the feedback enters a deeply nonlinear regime, and a small initial perturbation leads to the exponential divergence of adjacent trajectories, generating chaos with extreme sensitivity to initial conditions.

\subsection{Representative periodic and chaotic trajectories}

To further characterize the dynamics, we select representative parameter points corresponding to periodic and chaotic states and analyze their time-domain, frequency-domain, and phase-space behavior.

\begin{figure}[htbp]
    \centering
    \includegraphics[width=1\linewidth]{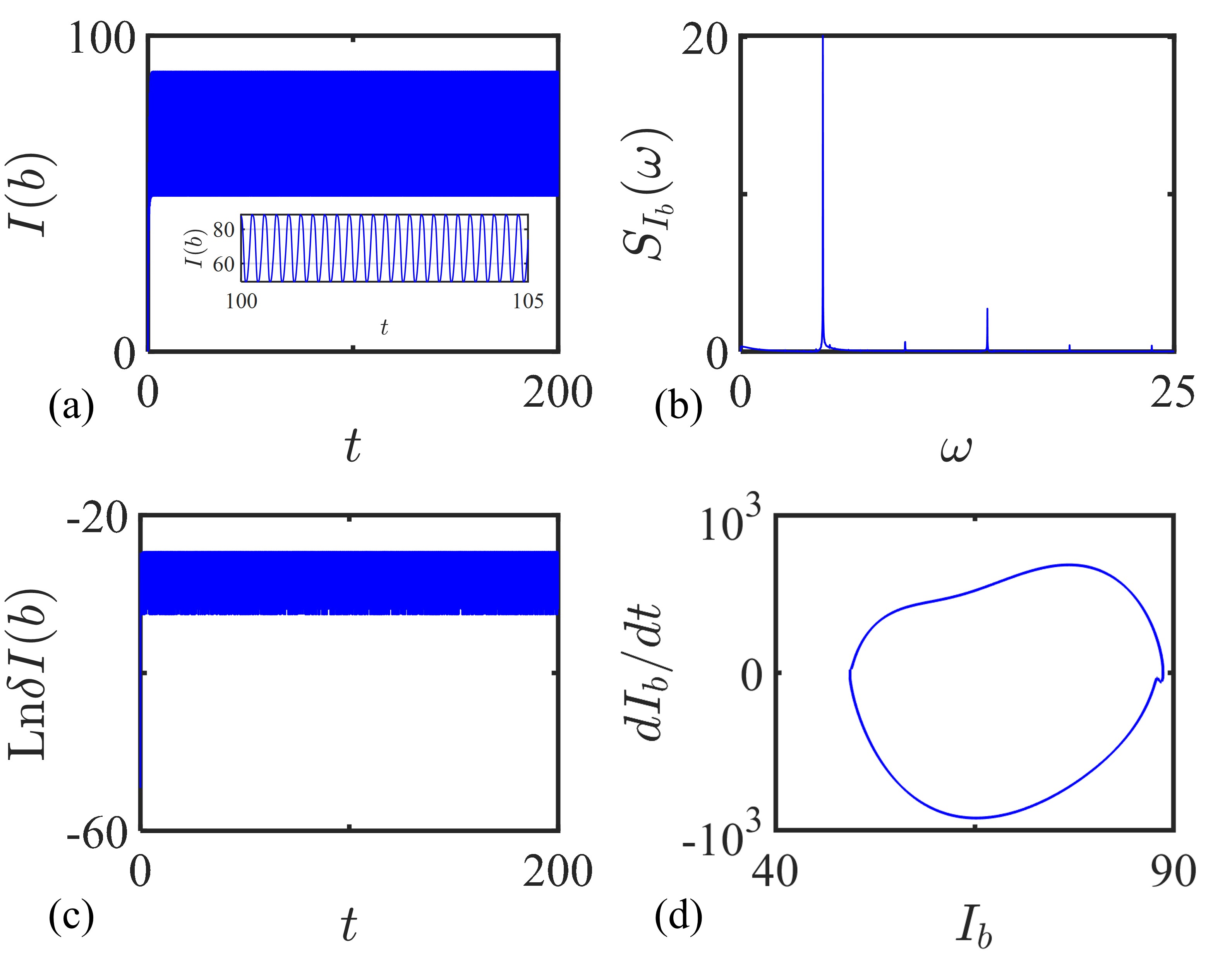}
    \caption{(a) Time evolution of the molecular vibration intensity \(I_b\). (b) Corresponding power spectrum \(\ln S_{I_b}(\omega)\). (c) Evolution of the logarithm of the perturbation \(\ln \delta I_b\). (d) Phase portrait in the \((dI_b/dt, I_b)\) plane. The parameters are \(J/2\pi = 0.38\,\mathrm{THz}\), \(g_c/2\pi = 7\,\mathrm{THz}\), \(\Delta_a/2\pi = 23\,\mathrm{THz}\), where the system exhibits typical periodic motion.}
    \label{fig:xiangtu1}
\end{figure}

Figure~\ref{fig:xiangtu1} displays the hallmarks of periodic motion: \(I_b\) undergoes regular oscillations with constant amplitude and frequency, and the waveform is strictly repetitive (Fig.~\ref{fig:xiangtu1}(a)). The corresponding spectrum consists of discrete sharp peaks without a continuous background (Fig.~\ref{fig:xiangtu1}(b)), reflecting the strict periodicity of the orbit. The perturbation logarithm \(\ln\delta I_b\) remains bounded with no persistent growth (Fig.~\ref{fig:xiangtu1}(c)), indicating that adjacent trajectories do not separate exponentially. The phase-space trajectory forms a closed limit cycle in the \((dI_b/dt, I_b)\) plane (Fig.~\ref{fig:xiangtu1}(d)), confirming the periodic nature of the motion.

\begin{figure}[htbp]
    \centering
    \includegraphics[width=1\linewidth]{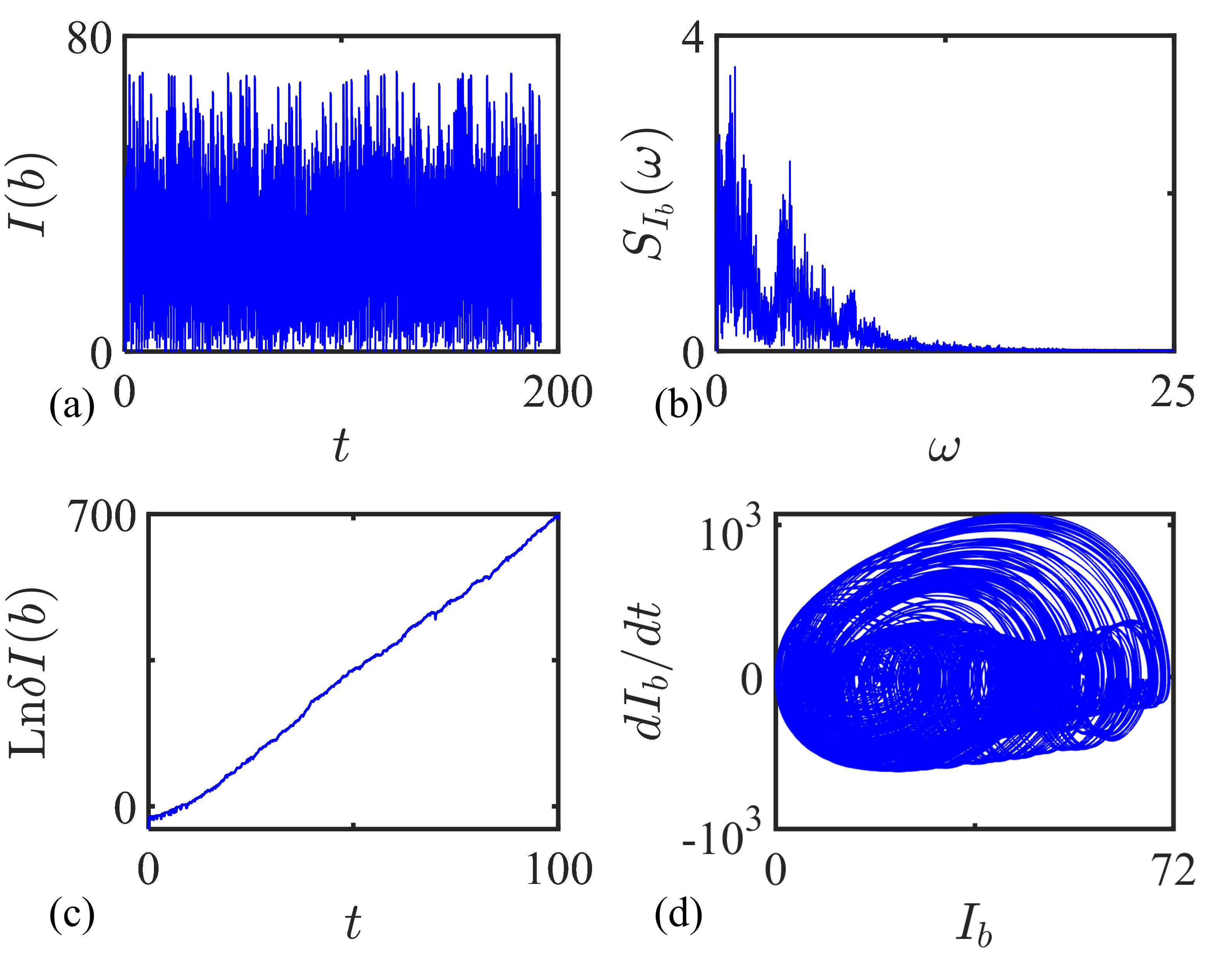}
    \caption{Chaotic dynamics for the same parameters as in Fig.~\ref{fig:xiangtu1}, except that the WGM detuning is changed to $\Delta_a/2\pi=45\,\mathrm{THz}$. The panels show (a) $I_b(t)$, (b) $S_{I_b}(\omega)$, (c) $\ln\delta I_b$, and (d) the phase portrait in the $(I_b,dI_b/dt)$ plane.}
    \label{fig:xiangtu2}
\end{figure}

In contrast, Fig.~\ref{fig:xiangtu2} illustrates the signatures of chaos. The time evolution of \(I_b\) (Fig.~\ref{fig:xiangtu2}(a)) shows pronounced aperiodic fluctuations with irregular amplitude variations and no repeating pattern—the defining feature of deterministic chaos. The corresponding spectrum (Fig.~\ref{fig:xiangtu2}(b)) becomes a continuous broadband structure, arising from the loss of periodicity. The perturbation logarithm \(\ln\delta I_b\) (Fig.~\ref{fig:xiangtu2}(c)) exhibits a clear upward trend, demonstrating the exponential divergence of initially infinitesimally close trajectories and thus extreme sensitivity to initial conditions. Physically, this originates from the amplification by the nonlinear terms \(2g_c b_r c_{i,r}\): within the chaotic regime, the plasmon intensity and the molecular displacement mutually drive each other, and a small perturbation is successively amplified through the optomechanical feedback loop. The phase-space trajectory (Fig.~\ref{fig:xiangtu2}(d)) displays the typical stretching-and-folding structure of a strange attractor, never closing within a finite region—the geometric fingerprint of deterministic chaos.

\section{Conclusion}\label{four}

We have theoretically established a hybrid molecular cavity optomechanical system, composed of a WGM microcavity coupled to a plasmonic nanocavity, as a versatile platform for room-temperature chaos. By computing the largest Lyapunov exponent and mapping phase diagrams in the spaces of plasmon–molecule coupling, plasmon–WGM coupling, and optical detuning, we have uncovered a complete route from self-sustained oscillation through a period-doubling cascade to chaos, and accurately delineated the parameter boundaries of each dynamical regime. The plasmon-vibration coupling \(g_c\) is the dominant nonlinear driving force, while the inter-cavity coupling \(J\) and the optical detuning \(\Delta_a\) provide flexible, cooperative control over the generation and evolution of chaos. Remarkably, even at moderate \(g_c\), isolated chaos windows can be opened simply by tuning \(\Delta_a\) and \(J\), demonstrating the high controllability of the system. This architecture exploits terahertz molecular vibrations to inherently suppress thermal noise, circumventing the stringent requirements for cryogenic temperatures or high driving power that limit conventional optomechanical systems. Our results lay a theoretical foundation for room-temperature, on-chip integrated low-threshold chaotic light sources and ultrahigh-speed physical random number generators, and provide crucial support for secure communication technologies based on optomechanical chaos. Future work may explore the role of quantum effects in regulating the chaotic dynamics of this system, as well as practical schemes for chaotic secure communication on this platform.

\section*{Acknowledgements}
This work was partly supported by the National Natural Science Foundation of China under Grant No. 12265022, the Inner Mongolia Natural Science Foundation under Grant No. 2025MS01005 and No. 2026MS0476, and Elite Revitalizing Inner Mongolia Program (2025TGL05).

\bibliography{ref}

@PREAMBLE{
 "\providecommand{\noopsort}[1]{}" 
 # "\providecommand{\singleletter}[1]{#1}%" 
}

@article{AllariaSynchronization01,
  title = {Synchronization of Homoclinic Chaos},
  author = {Allaria, E. and Arecchi, F. T. and Di Garbo, A. and Meucci, R.},
  journal = {Phys. Rev. Lett.},
  volume = {86},
  issue = {5},
  pages = {791--794},
  numpages = {0},
  year = {2001},
  month = {Jan},
  publisher = {American Physical Society},
  doi = {10.1103/PhysRevLett.86.791},
  url = {https://link.aps.org/doi/10.1103/PhysRevLett.86.791}
}

@article{WOS:000380760000001,
  title = {From quantum chaos and eigenstate thermalization to statistical mechanics and thermodynamics},
  author = {D'Alessio, Luca and Kafri, Yariv and Polkovnikov, Anatoli and Rigol, Marcos},
  journal = {Adv. Phys.},
  volume = {65},
  issue = {3},
  pages = {239-362},
  year = {2016},
  month = {May},
  publisher = {Taylor \& Francis},
  doi = {10.1080/00018732.2016.1198134},
  url = {https://doi.org/10.1080/00018732.2016.1198134}
}

@article{WOS:000245871200063,
  title = {Chaotic Quivering of Micron-Scaled On-Chip Resonators Excited by Centrifugal Optical Pressure},
  author = {Carmon, Tal and Cross, M. C. and Vahala, Kerry J.},
  journal = {Phys. Rev. Lett.},
  volume = {98},
  issue = {16},
  pages = {167203},
  numpages = {4},
  year = {2007},
  month = {Apr},
  publisher = {American Physical Society},
  doi = {10.1103/PhysRevLett.98.167203},
  url = {https://link.aps.org/doi/10.1103/PhysRevLett.98.167203}
}

@book{thompson2002nonlinear,
  title={Nonlinear dynamics and chaos},
  author={Thompson, John Michael Tutill and Stewart, H Bruce},
  year={2002},
  publisher={John Wiley \& Sons}
}

@article{ChenPhysical25,
  title = {Physical layer encryption based on digital chaos in THz wireless communication},
  author = {Chen, Qinghui and Peng, Cong and Zhao, Li and You, Weihao and Wen, Hong and Li, Jianshe},
  journal = {Opt. Fiber Technol.},
  volume = {90},
  issue = {},
  pages = {104118},
  year = {2025},
  month = {Jan},
  publisher = {Elsevier},
  doi = {10.1016/j.yofte.2024.104118},
  url = {https://www.sciencedirect.com/science/article/pii/S1068520024004632}
}

@article{ PengBroadband24,
author = {Yi-Bo Peng and Zhecheng Dai and Kai-Li Lin and Peng-Lei Wang and Zhijian Shen and Baile Chen and Fr\'{e}d\'{e}ric Grillot and Cheng Wang},
journal = {Opt. Lett.},
number = {11},
pages = {3142--3145},
publisher = {Optica Publishing Group},
title = {Broadband chaos of an interband cascade laser with a 6-GHz bandwidth},
volume = {49},
month = {Jun},
year = {2024},
url = {https://opg.optica.org/ol/abstract.cfm?URI=ol-49-11-3142},
doi = {10.1364/OL.525636},
}

@article{ LUPT15,
  title = {$\mathcal{P}\mathcal{T}$-Symmetry-Breaking Chaos in Optomechanics},
  author = {L\"u, Xin-You and Jing, Hui and Ma, Jin-Yong and Wu, Ying},
  journal = {Phys. Rev. Lett.},
  volume = {114},
  issue = {25},
  pages = {253601},
  numpages = {6},
  year = {2015},
  month = {Jun},
  publisher = {American Physical Society},
  doi = {10.1103/PhysRevLett.114.253601},
  url = {https://link.aps.org/doi/10.1103/PhysRevLett.114.253601}
}

@article{ MonifiOptomechanically16,
  title = {Optomechanically induced stochastic resonance and chaos transfer between optical fields},
  author={Monifi, Faraz and Zhang, Jing and {\"O}zdemir, {\c{S}}ahin Kaya and Peng, Bo and Liu, Yu-xi and Bo, Fang and Nori, Franco and Yang, Lan},
  journal = {Nat. Photon.},
  volume = {10},
  issue = {6},
  pages = {399-405},
  year = {2016},
  month = {Jun},
  publisher={Nature Publishing Group UK London},
  doi = {10.1038/nphoton.2016.73},
  url = {https://doi.org/10.1038/nphoton.2016.73}
}

@article{ Navarro-UrriosNonlinear17,
  title = {Nonlinear dynamics and chaos in an optomechanical beam},
  author={Navarro-Urrios, Daniel and Capuj, N{\'e}stor E and Colombano, Mart{\'\i}n F and Garc{\'\i}a, P David and Sledzinska, Marianna and Alzina, Francesc and Griol, Amadeu and Mart{\'\i}nez, Alejandro and Sotomayor-Torres, Clivia M},
  journal = {Nat. Commun.},
  volume = {8},
  issue = {1},
  pages = {14965},
  year = {2017},
  month = {Apr},
  publisher={Nature Publishing Group UK London},
  doi = {10.1038/ncomms14965},
  url = {https://doi.org/10.1038/ncomms14965}
}

@article{ ZhangNonreciprocal21,
  title = {Nonreciprocal chaos in a spinning optomechanical resonator},
  author = {Zhang, Deng-Wei and Zheng, Li-Li and You, Cai and Hu, Chang-Sheng and Wu, Ying and L\"u, Xin-You},
  journal = {Phys. Rev. A},
  volume = {104},
  issue = {3},
  pages = {033522},
  numpages = {7},
  year = {2021},
  month = {Sep},
  publisher = {American Physical Society},
  doi = {10.1103/PhysRevA.104.033522},
  url = {https://link.aps.org/doi/10.1103/PhysRevA.104.033522}
}

@article{ RoqueNonlinear20,
doi = {10.1088/1367-2630/ab6522},
url = {https://doi.org/10.1088/1367-2630/ab6522},
year = {2020},
month = {jan},
publisher = {IOP Publishing},
volume = {22},
number = {1},
pages = {013049},
author = {Roque, Thales Figueiredo and Marquardt, Florian and Yevtushenko, Oleg M},
title = {Nonlinear dynamics of weakly dissipative optomechanical systems},
journal = {New J. Phys.},
}

@article{ ZhangSqueezing-induced25,
  title = {Squeezing-induced nonreciprocal chaos in a cavity optomechanical system},
  author = {Zhang, Zhen-Yu and Qian, Yi-Bing and Sun, Lei and Huang, Shi-Tong and Hou, Bang-Pin and Tang, Lei},
  journal = {Phys. Lett. A},
  volume = {533},
  pages = {130222},
  year = {2025},
  publisher = {Elsevier},
  doi = {10.1016/j.physleta.2025.130222},
  url = {https://www.sciencedirect.com/science/article/pii/S0375960125000027}
}

@article{ZhuCavity23,
  title = {Cavity optomechanical chaos},
  author = {Gui-Lei Zhu and Chang-Sheng Hu and Ying Wu and Xin-You Lü},
  journal = {Fundam. Res.},
  volume = {3},
  issue = {1},
  pages = {63-74},
  year = {2023},
  doi = {10.1016/j.fmre.2022.07.012},
  url = {https://www.sciencedirect.com/science/article/pii/S2667325822003387}
}

@article{ LiuPhase-mediated19,
author = {Zeng-Xing Liu and Cai You and Bao Wang and Hao Xiong and Ying Wu},
journal = {Opt. Lett.},
number = {3},
pages = {507--510},
publisher = {Optica Publishing Group},
title = {Phase-mediated magnon chaos-order transition in cavity optomagnonics},
volume = {44},
month = {Feb},
year = {2019},
url = {https://opg.optica.org/ol/abstract.cfm?URI=ol-44-3-507},
doi = {10.1364/OL.44.000507},
}

@article{ LiNonlinear23,
  title = {Nonlinear self-sustaining dynamics in cavity magnomechanics},
  author = {Li, Wenlin and Cheng, Jiong and Gong, Wei-jiang and Li, Jie},
  journal = {Phys. Rev. A},
  volume = {108},
  issue = {3},
  pages = {033518},
  numpages = {11},
  year = {2023},
  month = {Sep},
  publisher = {American Physical Society},
  doi = {10.1103/PhysRevA.108.033518},
  url = {https://link.aps.org/doi/10.1103/PhysRevA.108.033518}
}

@article{ PengCavity24,
  title = {Cavity magnomechanical chaos},
  author = {Peng, Jiao and Liu, Zeng-Xing and Yu, Ya-Fei and Xiong, Hao},
  journal = {Phys. Rev. A},
  volume = {110},
  issue = {5},
  pages = {053704},
  numpages = {8},
  year = {2024},
  month = {Nov},
  publisher = {American Physical Society},
  doi = {10.1103/PhysRevA.110.053704},
  url = {https://link.aps.org/doi/10.1103/PhysRevA.110.053704}
}

@article{WangMagnon19,
  title = {Magnon chaos in $ PT $-symmetric cavity magnomechanics},
  author = {Wang, Mei and Zhang, Duo and Li, Xiang-Hu and Wu, Yu-Ying and Sun, Zhao-Yu},
  journal = {IEEE Photon. J.},
  volume = {11},
  issue = {3},
  pages = {1-8},
  year = {2019},
  month = {Jun},
  publisher = {IEEE},
  doi = {10.1109/JPHOT.2019.2911963},
  url = {https://doi.org/10.1109/JPHOT.2019.2911963}
}

@article{Xuchaos25,
author = {Hui Hui Xu and Feng Ze Cao and E Wu and Yong Hong Ma},
journal = {Opt. Express},
number = {24},
pages = {49996--50006},
publisher = {Optica Publishing Group},
title = {Chaos generation and control in optoelectromechanical systems},
volume = {33},
month = {Dec},
year = {2025},
url = {https://opg.optica.org/oe/abstract.cfm?URI=oe-33-24-49996},
doi = {10.1364/OE.577725},
}

@article{WangControllable16,
  title = {Controllable chaos in hybrid electro-optomechanical systems},
  author={Wang, Mei and L{\"u}, Xin-You and Ma, Jin-Yong and Xiong, Hao and Si, Liu-Gang and Wu, Ying},
  journal = {Sci. Rep.},
  volume={6},
  number={1},
  pages={22705},
  year = {2016},
  month = {Mar},
  publisher={Nature Publishing Group UK London},
  doi = {10.1038/srep22705},
  url = {https://doi.org/10.1038/srep22705}
}

@article{BakemeierRoute15,
  title = {Route to Chaos in Optomechanics},
  author = {Bakemeier, L. and Alvermann, A. and Fehske, H.},
  journal = {Phys. Rev. Lett.},
  volume = {114},
  issue = {1},
  pages = {013601},
  numpages = {5},
  year = {2015},
  month = {Jan},
  publisher = {American Physical Society},
  doi = {10.1103/PhysRevLett.114.013601},
  url = {https://link.aps.org/doi/10.1103/PhysRevLett.114.013601}
}

@article{ LiRoutes24,
  title = {Routes to chaos in the balanced two-photon Dicke model with qubit dissipation},
  author = {Li, Jiahui and Chesi, Stefano},
  journal = {Phys. Rev. A},
  volume = {109},
  issue = {5},
  pages = {053702},
  numpages = {9},
  year = {2024},
  month = {May},
  publisher = {American Physical Society},
  doi = {10.1103/PhysRevA.109.053702},
  url = {https://link.aps.org/doi/10.1103/PhysRevA.109.053702}
}

@article{ HalefRoute25,
  title = {Route to Hyperchaos in Quadratic Optomechanics},
  author = {Halef, Lina and Shomroni, Itay},
  journal = {Phys. Rev. Lett.},
  volume = {135},
  issue = {25},
  pages = {257201},
  numpages = {10},
  year = {2025},
  month = {Dec},
  publisher = {American Physical Society},
  doi = {10.1103/rv1f-x73d},
  url = {https://link.aps.org/doi/10.1103/rv1f-x73d}
}

@article{ GenesRobust08,
  title = {Robust entanglement of a micromechanical resonator with output optical fields},
  author = {Genes, C. and Mari, A. and Tombesi, P. and Vitali, D.},
  journal = {Phys. Rev. A},
  volume = {78},
  issue = {3},
  pages = {032316},
  numpages = {14},
  year = {2008},
  month = {Sep},
  publisher = {American Physical Society},
  doi = {10.1103/PhysRevA.78.032316},
  url = {https://link.aps.org/doi/10.1103/PhysRevA.78.032316}
}

@article{ WangReservoir-Engineered13,
  title = {Reservoir-Engineered Entanglement in Optomechanical Systems},
  author = {Wang, Ying-Dan and Clerk, Aashish A.},
  journal = {Phys. Rev. Lett.},
  volume = {110},
  issue = {25},
  pages = {253601},
  numpages = {5},
  year = {2013},
  month = {Jun},
  publisher = {American Physical Society},
  doi = {10.1103/PhysRevLett.110.253601},
  url = {https://link.aps.org/doi/10.1103/PhysRevLett.110.253601}
}

@article{ LiBoosting24,
  title = {Boosting Light-Matter Interactions in Plasmonic Nanogaps},
  author = {Li, Yang and Chen, Wen and He, Xiaobo and Shi, Junjun and Cui, Ximin and Sun, Jiawei and Xu, Hongxing},
  journal = {Adv. Mater.},
  volume = {36},
  issue = {49},
  pages = {2405186},
  year = {2024},
  month = {Dec},
  publisher = {Wiley Online Library},
  doi = {10.1002/adma.202405186},
  url = {https://advanced.onlinelibrary.wiley.com/doi/abs/10.1002/adma.202405186}
}

@article{EstebanMolecular22,
  title = {Molecular Optomechanics Approach to Surface-Enhanced Raman Scattering},
  author = {Esteban, Ruben and Baumberg, Jeremy J. and Aizpurua, Javier},
  journal = {Acc. Chem. Res.},
  volume = {55},
  issue = {14},
  pages = {1889-1899},
  numpages = {11},
  year = {2022},
  month = {Jul},
  publisher = {ACS Publications},
  doi = {10.1021/acs.accounts.1c00759},
  url = {https://doi.org/10.1021/acs.accounts.1c00759}
}

@article{ MaTwo-Polariton26,
  title = {Two-Polariton Blockade via Ultrastrong Light-Matter Coupling},
  author = {Ma, Ting-Ting and Tang, Jian and Zuo, Yun-Lan and Huang, Ran and Miranowicz, Adam and Nori, Franco and Jing, Hui},
  journal = {Phys. Rev. Lett.},
  volume = {136},
  issue = {3},
  pages = {033601},
  numpages = {9},
  year = {2026},
  month = {Jan},
  publisher = {American Physical Society},
  doi = {10.1103/nfz3-txyt},
  url = {https://link.aps.org/doi/10.1103/nfz3-txyt}
}

@article{RoelliMolecular16,
  title = {Molecular cavity optomechanics as a theory of plasmon-enhanced Raman scattering},
  author = {Roelli, Philippe and Galland, Christophe and Piro, Nicolas and Kippenberg, Tobias J.},
  journal = {Nat. Nanotechnol.},
  volume = {11},
  issue = {2},
  pages = {164-169},
  year = {2016},
  month = {Feb},
  publisher = {Nature Publishing Group},
  doi = {10.1038/nnano.2015.264},
  url = {https://doi.org/10.1038/nnano.2015.264}
}

@article{SchmidtQuantum16,
  title = {Quantum Mechanical Description of Raman Scattering from Molecules in Plasmonic Cavities},
  author = {Schmidt, Mikolaj K. and Esteban, Ruben and Gonzalez-Tudela, Alejandro and Giedke, Geza and Aizpurua, Javier},
  journal = {ACS Nano},
  volume = {10},
  issue = {6},
  pages = {6291-6298},
  year = {2016},
  month = {Jun},
  publisher = {American Chemical Society},
  doi = {10.1021/acsnano.6b02484},
  url = {https://doi.org/10.1021/acsnano.6b02484}
}

@article{SchmidtLinking17,
  title = {Linking classical and molecular optomechanics descriptions of SERS},
  author = {Schmidt, Miko{\l}aj K and Esteban, Ruben and Benz, Felix and Baumberg, Jeremy J and Aizpurua, Javier},
  journal = {Faraday Discuss.},
  volume = {205},
  issue = {0},
  pages = {31-65},
  year = {2017},
  month = {Jun},
  publisher = {The Royal Society of Chemistry},
  doi = {10.1039/C7FD00145B},
  url = {http://dx.doi.org/10.1039/C7FD00145B}
}

@article{ LombardiPulsed18,
  title = {Pulsed Molecular Optomechanics in Plasmonic Nanocavities: From Nonlinear Vibrational Instabilities to Bond-Breaking},
  author = {Lombardi, Anna and Schmidt, Miko\l{}aj K. and Weller, Lee and Deacon, William M. and Benz, Felix and de Nijs, Bart and Aizpurua, Javier and Baumberg, Jeremy J.},
  journal = {Phys. Rev. X},
  volume = {8},
  issue = {1},
  pages = {011016},
  numpages = {17},
  year = {2018},
  month = {Feb},
  publisher = {American Physical Society},
  doi = {10.1103/PhysRevX.8.011016},
  url = {https://link.aps.org/doi/10.1103/PhysRevX.8.011016}
}

@article{ AndersonTwo-Color18,
  title = {Two-Color Pump-Probe Measurement of Photonic Quantum Correlations Mediated by a Single Phonon},
  author = {Anderson, Mitchell D. and Tarrago Velez, Santiago and Seibold, Kilian and Flayac, Hugo and Savona, Vincenzo and Sangouard, Nicolas and Galland, Christophe},
  journal = {Phys. Rev. Lett.},
  volume = {120},
  issue = {23},
  pages = {233601},
  numpages = {5},
  year = {2018},
  month = {Jun},
  publisher = {American Physical Society},
  doi = {10.1103/PhysRevLett.120.233601},
  url = {https://link.aps.org/doi/10.1103/PhysRevLett.120.233601}
}

@article{ BellCoherent25,
  title = {Coherent Dynamics of Molecular Vibrations in Single Plasmonic Nanogaps},
  author = {Bell, Fiona and Jakob, Lukas and Todd, Caleb and Lohia, Ishaan and Roh, Yeeun and Arul, Rakesh and Baumberg, Jeremy J.},
  journal = {Phys. Rev. Lett.},
  volume = {135},
  issue = {7},
  pages = {076901},
  numpages = {7},
  year = {2025},
  month = {Aug},
  publisher = {American Physical Society},
  doi = {10.1103/txdw-nqvn},
  url = {https://link.aps.org/doi/10.1103/txdw-nqvn}
}

@article{ KimUltrasmall16,
  title={Ultrasmall nanoparticles induce ferroptosis in nutrient-deprived cancer cells and suppress tumour growth},
  author={Kim, Sung Eun and Zhang, Li and Ma, Kai and Riegman, Michelle and Chen, Feng and Ingold, Irina and Conrad, Marcus and Turker, Melik Ziya and Gao, Minghui and Jiang, Xuejun and Monette, Sebastien and Pauliah, Mohan and Gonen, Mithat and Zanzonico, Pat and Quinn, Thomas and Wiesner, Ulrich and Bradbury, Michelle S. and Overholtzer, Michael},
  journal={Nat. Nanotechnol.},
  volume={11},
  number={11},
  pages={977--985},
  year={2016},
  month={11},
  doi={10.1038/nnano.2016.164},
  issn={1748-3395},
  url={https://doi.org/10.1038/nnano.2016.164}
}

@article{ HuangCollective24,
  title = {Collective quantum entanglement in molecular cavity optomechanics},
  author = {Huang, Jian and Lei, Dangyuan and Agarwal, Girish S. and Zhang, Zhedong},
  journal = {Phys. Rev. B},
  volume = {110},
  issue = {18},
  pages = {184306},
  numpages = {10},
  year = {2024},
  month = {Nov},
  publisher = {American Physical Society},
  doi = {10.1103/PhysRevB.110.184306},
  url = {https://link.aps.org/doi/10.1103/PhysRevB.110.184306}
}

@article{YinMolecular-Optomechanical26,
  title = {Molecular-Optomechanical Phonon Laser},
  author = {Yin, Bin and Wang, Jie and Zhang, Qian and Wang, Deng and Lu, Tian-Xiang and Jing, Hui},
  journal = {Laser \& Photonics Reviews},
  pages = {e71191},
  year = {2026},
  publisher = {Wiley Online Library},
  doi = {10.1002/lpor.71191},
  url = {https://onlinelibrary.wiley.com/doi/abs/10.1002/lpor.71191}
}

@article{ BerinyuyQuantum26,
  title = {Quantum correlations in molecular cavity optomechanics},
  author = {Berinyuy, E. Kongkui and Massembele, D.R. Kenigoule and Djorwé, P. and Altuijri, R. and Abdel-Khalek, S. and Abdel-Aty, A.-H. and Engo, S.G. Nana},
  journal = {Chaos, Solitons \& Fractals},
  volume = {205},
  issue = {},
  pages = {117820},
  numpages = {},
  year = {2026},
  month = {},
  publisher = {Elsevier},
  doi = {10.1016/j.chaos.2025.117820},
  url = {https://www.sciencedirect.com/science/article/pii/S096007792501834X}
}

@article{ Loirette-PelousAddressing26,
  title = {Addressing Intramolecular Vibrational Redistribution in a Single Molecule through Pump and Probe Surface-Enhanced Vibrational Spectroscopy},
  author = {Loirette-Pelous, Aurelian and Boto, Roberto A. and Aizpurua, Javier and Esteban, Ruben},
  journal = {ACS Photonics},
  volume = {},
  issue = {},
  pages = {},
  year = {2026},
  month = {May},
  publisher = {American Chemical Society},
  doi = {10.1021/acsphotonics.6c00030},
  url = {https://doi.org/10.1021/acsphotonics.6c00030}
}

@article{ BarzanjehStationary19,
  title = {Stationary entangled radiation from micromechanical motion},
  author = {Barzanjeh, S. and Redchenko, E. S. and Peruzzo, M. and Wulf, M. and Lewis, D. P. and Arnold, G. and Fink, J. M.},
  journal = {Nature},
  volume = {570},
  issue = {7762},
  pages = {480--483},
  year = {2019},
  month = {Jun},
  publisher = {Springer Nature},
  doi = {10.1038/s41586-019-1320-2},
  url = {https://doi.org/10.1038/s41586-019-1320-2}
}

@article{YuStrong26,
  title = {Strong Molecule-Light Entanglement with Molecular Cavity Optomechanics},
  author = {Yu, Hong-Yun and Jiao, Ya-Feng and Wang, Jie and Li, Feng and Yin, Bin and Liu, Qi-Rui and Jiang, Tian and Jing, Hui and Wei, Ke},
  journal = {Phys. Rev. Lett.},
  volume = {136},
  issue = {1},
  pages = {013602},
  numpages = {7},
  year = {2026},
  month = {Jan},
  publisher = {American Physical Society},
  doi = {10.1103/4z8v-f6s5},
  url = {https://link.aps.org/doi/10.1103/4z8v-f6s5}
}

@article{BenzSingle-molecule16,
author = {Felix Benz  and Mikolaj K. Schmidt  and Alexander Dreismann  and Rohit Chikkaraddy  and Yao Zhang  and Angela Demetriadou  and Cloudy Carnegie  and Hamid Ohadi  and Bart de Nijs  and Ruben Esteban  and Javier Aizpurua  and Jeremy J. Baumberg },
title = {Single-molecule optomechanics in “picocavities”},
journal = {Science},
volume = {354},
number = {6313},
pages = {726-729},
year = {2016},
doi = {10.1126/science.aah5243},
URL = {https://www.science.org/doi/abs/10.1126/science.aah5243},
}

@article{ BaumbergPicocavities22,
  title = {Picocavities: a Primer},
  author = {Baumberg, Jeremy J.},
  journal = {Nano Lett.},
  volume = {22},
  issue = {14},
  pages = {5859--5865},
  year = {2022},
  month = {Jul},
  publisher = {American Chemical Society},
  doi = {10.1021/acs.nanolett.2c01695},
  url = {https://doi.org/10.1021/acs.nanolett.2c01695}
}

@article{ BaumbergExtreme19,
  title = {Extreme nanophotonics from ultrathin metallic gaps},
  author = {Baumberg, Jeremy J. and Aizpurua, Javier and Mikkelsen, Maiken H. and Smith, David R.},
  journal = {Nat. Mater.},
  volume = {18},
  issue = {7},
  pages = {668--678},
  year = {2019},
  month = {Jul},
  doi = {10.1038/s41563-019-0290-y},
  url = {https://doi.org/10.1038/s41563-019-0290-y}
}

@article{ TangRobust26,
  title = {Robust photon blockade with hybrid molecular optomechanics},
  author = {Tang, Jian and Li, Baijun and Yin, Bin and Lu, Tian-Xiang and Huang, Ran and Nori, Franco and Jing, Hui},
  journal = {npj Quantum Inf.},
  volume = {12},
  issue = {1},
  pages = {78},
  year = {2026},
  month = {Mar},
  publisher = {Springer Nature},
  doi = {10.1038/s41534-026-01220-3},
  url = {https://doi.org/10.1038/s41534-026-01220-3}
}

@article{ YinMolecular25,
  title = {Molecular optomechanically induced transparency},
  author = {Yin, Bin and Wang, Jie and Peng, Mei-Yu and Zhang, Qian and Wang, Deng and Lu, Tian-Xiang and Wei, Ke and Jing, Hui},
  journal = {Phys. Rev. A},
  volume = {111},
  issue = {4},
  pages = {043507},
  numpages = {12},
  year = {2025},
  month = {Apr},
  publisher = {American Physical Society},
  doi = {10.1103/PhysRevA.111.043507},
  url = {https://link.aps.org/doi/10.1103/PhysRevA.111.043507}
}

@article{ NowoczynUniversal26,
  title = {Universal quantum melting of quasiperiodic attractors in driven-dissipative cavities},
  author = {Nowoczyn, Caroline and Mathey, Ludwig and Seibold, Kilian},
  journal = {Phys. Rev. A},
  volume = {113},
  issue = {5},
  pages = {052208},
  numpages = {13},
  year = {2026},
  month = {May},
  publisher = {American Physical Society},
  doi = {10.1103/pdnh-1yxs},
  url = {https://link.aps.org/doi/10.1103/pdnh-1yxs}
}

@article{escot2020brief,
  title = {A brief methodological note on chaos theory and its recent applications based on new computer resources},
  author = {Escot Mangas, Lorenzo and Sandubete Gal{\'a}n, Julio E},
  journal = {Energeia},
  volume = {VII},
  issue = {1},
  pages = {53--64},
  year = {2020},
  month = {},
  publisher = {Sociedad Ibero-Americana de Metodología Económica},
  doi = {},
  url = {https://hdl.handle.net/20.500.14352/8469}
}

@article{benettin1980lyapunov,
  title={Lyapunov characteristic exponents for smooth dynamical systems and for Hamiltonian systems; a method for computing all of them. Part 1: Theory},
  author={Benettin, Giancarlo and Galgani, Luigi and Giorgilli, Antonio and Strelcyn, Jean-Marie},
  journal={Meccanica},
  volume={15},
  number={1},
  pages={9--20},
  year={1980},
  month={3},
  publisher={Springer},
  issn={1572-9648},
  doi={10.1007/BF02128236},
  url={https://doi.org/10.1007/BF02128236}
}

@article{datseris2018dynamicalsystems,
  title = {DynamicalSystems. jl: A Julia software library for chaos and nonlinear dynamics},
  author = {Datseris, George},
  journal = {J. Open Source Softw.},
  volume = {3},
  issue = {23},
  pages = {598},
  year = {2018},
  month = {Mar},
  publisher = {The Open Journal},
  doi = {10.21105/joss.00598},
  url = {https://doi.org/10.21105/joss.00598}
}

@book{baier1991chaotic,
  title={A chaotic hierarchy},
  author={Baier, Gerold and Klein, Michael},
  year={1991},
  publisher={World Scientific}
}

@article{ JakobGiant23,
  title = {Giant optomechanical spring effect in plasmonic nano- and picocavities probed by surface-enhanced Raman scattering},
  author = {Jakob, Lukas A. and Deacon, William M. and Zhang, Yuan and de Nijs, Bart and Pavlenko, Elena and Hu, Shu and Carnegie, Cloudy and Neuman, Tomas and Esteban, Ruben and Aizpurua, Javier and Baumberg, Jeremy J.},
  journal = {Nat. Commun.},
  volume = {14},
  issue = {1},
  pages = {3291},
  year = {2023},
  month = {Jun},
  doi = {10.1038/s41467-023-38124-1},
  url = {https://doi.org/10.1038/s41467-023-38124-1}
}
\end{document}